\documentclass[paper]{JHEP3}
\usepackage{yfonts}[1988/10/03]
\newcommand {\beq}{\begin{equation}}
\newcommand {\eeq}{\end{equation}}
\newcommand {\beqa}{\begin{eqnarray}}
\newcommand {\eeqa}{\end{eqnarray}}
\newcommand {\n}{\nonumber \\}
\newcommand {\del}{\partial}
\newcommand{\g}{\textfrak{g}}
\newcommand{\R}{\textfrak{R}}

\title{Density-Metric Unimodular Gravity: vacuum spherical symmetry} 
\author{Amir H. Abbassi\\
Department of Physics, School of Sciences, Tarbiat Modares
University,\\ P.O.Box 14155-4838, Tehran, Iran.\\
E-mail:\email{ahabbasi@modares.ac.ir}}
\author{Amir M. Abbassi\\
Depatment of Physics, University of Tehran,\\ P.O.Box 14155-6455, Tehran, Iran.\\
E-mail:\email{amabasi@khayam.ut.ac.ir}}
\abstract{We analyze an alternative theory of gravity characterized by metrics
that are tensor density of rank $(0,2)$ and weight $-\frac12$. The metric compatibility condition is supposed to hold. The simplest expression for the action of gravitational field is used. Taking the metric and trace of connections as dynamical variables, the field equations in the absence of matter and other kinds of sources are derived. The solutions of these equations are obtained for the case of vacuum static spherical symmetric spacetime. The null geodesics and advance of perihelion of ellipses  are discussed. We confirm a subclass of solutions are regular for $r>0$ and there is no event horizon while it is singular at $r=0$. }
\keywords{Unimodular gravity,Modified gravity}
\preprint{***}
\begin{document}

\section{Introduction}
Unimodular theory of relativity was an alternative theory of gravity considered by Einstein in 1919. It turned out that it is equivalent to general relativity with the cosmological constant appearing as integration constant.
In unimodular theory and its modified versions the determinant of the metric does not serve as a dynamical variable and this imposes a specific constraint on the variations of the metric\cite{Ein1919,Fink71,Fink01,Bij,Wein89,Alv}.
The reason for this very feature of the theory is obscure. It is being considered as the substantial essence of the system and often is not invoked by any decisive evidence or physical interpretation.
It will be more appropriate if this feature be the emergent of the theory and attributes a natural property of it. This paper is an attempt to fulfill this task. What should be noticed in advance is that in special relativity all the Jacobians of the propounded transformations are identically equal to one. Therefore all tensors including the metrics may be considered as tensor densities with arbitrary weight. This rises the question that in passing toward general relativity which one should be kept to be dealt with, metric or density metric? Certainly in general relativity where there is not such a constraint on the Jacobians, consideration of a tensor density as metric will bring significant change in the obtained equations.
As the starting point let us propose that the metric should be considered as a symmetric tensor density of rank $(0,2)$ and weight of $-\frac12$, denoting it by $\g_{\mu\nu}$. This is chosen and accepted so that the determinant of $\g_{\mu\nu}$ becomes a scalar. Without lose of generality this scalar may be normalized to one.
We should distinctly emphasize that this proposal for the role of the metric is novel in our approach and has no acquaintance in the literature.
The metric compatibility condition is supposed to hold for the new metric too. In short we may summarize:\\
\bigskip
\beqa  
 & & \g_{\mu\nu}\;,\;\;{\mbox{ symmetric tensor density of weight} -\frac12} \label{eq:1}\\
 & & |\g|=1  \label{eq:2}\\ 
 & & \nabla_\lambda \g_{\mu\nu}=0 \label{eq:3}
 \eeqa
 The inverse of the metric  $\g^{\mu\nu}$ is a density tensor of rank (2,0) with the weight $+\frac 12$ which satisfies the following 
 \beq \label{eq:4}
 \g^{\mu\nu} \g_{\nu\lambda}=\delta^\mu_\lambda \;\;\;\; ,\;\;\;
 \nabla_\lambda \g^{\mu\nu}=0
 \eeq
 The metric compatibility condition Eq.(\ref{eq:3}) gives 
 \beq \label{eq:5}
 \del_\lambda \g_{\mu\nu}-\Gamma^\rho_{\lambda\mu}\g_{\rho\nu}
 -\Gamma^\rho_{\lambda\nu}\g_{\mu\rho}+\frac {1}{2}\Gamma^\rho_{\rho\lambda}\g_{\mu\nu}=0
 \eeq
Rewriting Eq.(\ref{eq:5}) by a cyclic permutation of the indices makes:
\beq \label{eq:6}
\del_\nu \g_{\lambda\mu}-\Gamma^\rho_{\nu\lambda}\g_{\rho\mu}
-\Gamma^\rho_{\nu\mu}\g_{\rho\lambda}+\frac12\Gamma^\rho_{\rho\nu}\g_{\lambda\mu}=0
\eeq
\beq \label{eq:7}
\del_\mu\g_{\nu\lambda}-\Gamma^\rho_{\mu\nu}\g_{\rho\lambda}
-\Gamma^\rho_{\mu\lambda}\g_{\nu\rho}+\frac12\Gamma^\rho_{\rho\mu}\g_{\nu\lambda}=0
\eeq
Adding Eqs.~(\ref{eq:5}) and (\ref{eq:6}) and subtracting Eq.~(\ref{eq:7}) gives
\beq  \label{eq:8}
\del_\lambda\g_{\mu\nu}+\del_\nu\g_{\lambda\mu}-\del_\mu\g_{\nu\lambda}
-2\Gamma^\rho_{\lambda\nu}\g_{\rho\mu}+\frac12(\Gamma^\rho_{\rho\lambda}
\g_{\mu\nu}+\Gamma^\rho_{\rho\nu}\g_{\lambda\mu}-\Gamma^\rho_{\rho\mu}
\g_{\nu\lambda})=0
\eeq
Multiplying Eq.(\ref{eq:8}) by $\g^{\nu\kappa}$ leads to
\beq \label{eq:9}
\Gamma^\kappa_{\lambda\nu}=\frac12\g^{\kappa\mu}(\del_\lambda\g_{\mu\nu}
+\del_\nu\g_{\lambda\mu}-\del_\mu\g_{\nu\lambda})+\frac14(\Gamma^\rho_
{\rho\lambda}\delta^\kappa_\nu+\Gamma^\rho_{\rho\nu}\delta^\kappa_\lambda
-\Gamma^\rho_{\rho\mu}\g^{\mu\kappa}\g_{\nu\lambda})
\eeq
Eq.(\ref{eq:9}) does not fix all the components of the connection in terms of the metric and its derivatives, the components of the trace of connection $\Gamma^\rho_{\rho\lambda}$ remain undetermined. Since the determinant of $\g_{\mu\nu}$ is taken to be one this automatically yields the following constraint on the variations of $\g_{\mu\nu}$
\beq \label{eq:10}
\g^{\mu\nu}\delta\g_{\mu\nu}=0
\eeq
which is a required condition for the unimodular relativity. Using the connection (\ref{eq:9}), we may define the Riemann curvature and Ricci tensors respectively by:
\beqa 
& \R^\rho_{\sigma\mu\nu}= &\del_\nu\Gamma^\rho_{\mu\sigma}-\del_\mu\Gamma
^\rho_{\nu\sigma}+\Gamma^\rho_{\nu\lambda}\Gamma^\lambda_{\mu\sigma}-
\Gamma^\rho_{\mu\lambda}\Gamma^\lambda_{\nu\sigma} \label{eq:11}\\
& \R_{\sigma\nu}= &\del_\nu\Gamma^\rho_{\rho\sigma}-\del_\rho\Gamma^
\rho_{\sigma\nu}+\Gamma^\rho_{\nu\lambda}\Gamma^\lambda_{\rho\sigma}
-\Gamma^\rho_{\rho\lambda}\Gamma^\lambda_{\nu\sigma}. \label{eq:12}
\eeqa
$\R^\rho_{\sigma\mu\nu}$ and $\R_{\sigma\nu}$ both are regular tensors
i.e. are tensor densities of weight zero.\\
Now a scalar density of weight $+\frac 12$ may be constructed by using Ricci tensor (\ref{eq:12}) and inverse metric $\g^{\mu\nu}$,
\beq \label{eq:13}
\R=\g^{\mu\nu}\R_{\mu\nu}
\eeq
Since $d^4x$ is a scalar density of weight -1 , it is no longer possible to use determinant of $\g_{\mu\nu}$ to construct a scalar. 

\section{Action}
Let us consider gravitational fields in the absence of matter and other kinds of sources. \R $\;$ and $d^4x$ both are scalar densities of weight $\frac12$ and $-1$ respectively. Therefore $\R^2d^4x$ is a scalar. We may define the action as;
\beq \label{eq:14}
I=\int{\kappa \;\R^2 d^4x}
\eeq
where $\kappa$ is a proper constant. It is a functional of the metric, its first and second derivatives, and the unspecified trace of the connections $\Gamma^\rho_{\rho\lambda}$ and its first derivatives. It is somehow similar to Palatini approach where both metric and the connection are considered as independent dynamical variables. Here our dynamical variables are $\g^{\mu\nu}$ and $\Gamma^\rho_{\rho\lambda}$. The variations of the action with respect to the first leads to
\beq \label{eq:15}
\delta I=\int{ 2\kappa\;\R(\R_{\mu\nu}\delta\g^{\mu\nu}+\g^{\mu\nu}\delta\R_{\mu\nu})d^4x} =0
\eeq
while its variations with respect to the second gives
\beq \label{eq:16}
\delta I=\int{2\kappa\; \R\;\g^{\mu\nu}\delta\R_{\mu\nu}d^4x}=0
\eeq
Generally we have 
\beq \label{eq:17}
\delta \R_{\mu\nu}=\nabla_\rho (\delta\Gamma^\rho_{\mu\nu})-\nabla_\nu
(\delta\Gamma^\rho_{\mu\rho})
\eeq
The variation of $\Gamma^\kappa_{\mu\nu}$ with respect to $\Gamma^\rho_
{\rho\lambda}$ is
\beq \label{eq:18}
\delta\Gamma^\kappa_{\mu\nu}=\frac14(\delta^\lambda_\mu\delta^\kappa_\nu
+\delta^\lambda_\nu\delta^\kappa_\mu-\g^{\lambda\kappa}\g_{\mu\nu})
\delta\Gamma^\rho_{\rho\lambda}
\eeq
Inserting Eq.~(\ref{eq:18}) in Eq.~(\ref{eq:17}) and then Eq.~(\ref{eq:17}) 
in Eq.~(\ref{eq:16}) lead to the following equation:
\beq \label{eq:19}
\nabla_\lambda \R=0
\eeq
The variation of the action with respect to metric and applying the unimodular condition Eq.(
\ref{eq:10}) by the method of Lagrange undetermined
multipliers and inserting field equation (\ref{eq:19}) leads to
\beq \label{eq:20}
\R_{\mu\nu}-\frac14\g_{\mu\nu}\R=0.
\eeq
It is worth to notice that Eqs.(\ref{eq:19}) and (\ref{eq:20}) are consistent with the Bianchi identity. Field equation (\ref{eq:20}) is traceless and since \R$\;$ is a scalar density so Eq.(\ref{eq:19})
is not a trivial relation.\\
For the next we are going to find the analog of Schwarzschild spacetime
in this alternative theory. That is to find the solution of Eqs.(\ref{eq:19}) and (\ref{eq:20}) for a spherically symmetric spacetime.

\section{Spherical Symmetry}
In Schwarzschild solution the starting ansatz is that there exits a spherical coordinate system $x^\mu=(t,r,\theta\phi)$ in which the line
element has the form
\beq \label{eq:21}
ds^2=B(r)dt^2-A(r)dr^2-r^2(d\theta^2+sin^2\theta d\phi^2).
\eeq
If we define $\g_{\mu\nu}=\frac{g_{\mu\nu}}{\root 4\of{|\g|}}$ then it will have the desired property. So the components of the $\g_{\mu\nu}$
are:
\beqa \label{eq:22}
& \g_{tt} &=-\frac{B^\frac 34}{A^\frac14 r\sin^\frac12\theta}, \hspace{1cm}
\g_{rr}=\frac{A^\frac34}{B^\frac14 r\sin^\frac12\theta},\n 
& \g_{\theta\theta} &=\frac{r}{(AB)^\frac14\sin^\frac12\theta}, \hspace{9mm}
\g_{\phi\phi}=\frac{r\sin^\frac32\theta}{(AB)^\frac14}.
\eeqa
and the components of the inverse metric $\g^{\mu\nu}$ are :
\beqa \label{eq:23}
& \g^{tt} &=-{A^\frac14 r\sin^\frac12\theta\over B^\frac34}, \hspace{1cm}
\g^{rr}={B^\frac14 r\sin^\frac12\theta\over A^\frac34}, \n
& \g^{\theta\theta} &={(AB)^\frac14\sin^\frac12\theta\over r}, \hspace{9mm}
\g^{\phi\phi}={(AB)^\frac14\over r\sin^\frac32\theta}.
\eeqa
The nonzero components of the connection by using Eqs.(\ref{eq:9}),(\ref{eq:22})
and(\ref{eq:23}) are as follows;
\beqa \label{eq:24}
& & \Gamma^t_{tt}=\frac14\Gamma^\rho_{\rho t},\;\;\Gamma^t_{tr}=\frac38
\frac{B^\prime}B-\frac18\frac{A^\prime}{A}-\frac1{2r}+\frac14\Gamma^\rho
_{\rho r},\;\;\Gamma^t_{t\theta}=-\frac14\cot\theta+\frac14\Gamma^\rho_
{\rho\theta}, \n
& & \Gamma^t_{t\phi}=\frac14\Gamma^\rho_{\rho\phi},\;\;\Gamma^t_{rr}=
\frac14\frac AB\Gamma^\rho_{\rho t},\;\;\Gamma^t_{\theta\theta}=
\frac14\frac{r^2}{B}\Gamma^\rho_{\rho t},\;\;\Gamma^t_{\phi\phi}=
\sin^2\theta\Gamma^t_{\theta\theta}, \n
& & \Gamma^r_{tt}=\frac38\frac{B^\prime}{A}-\frac18\frac{BA^\prime}{A^2}
-\frac{B}{2rA}+\frac14\frac BA\Gamma^\rho_{\rho r},\;\;\Gamma^r_{rt}=
\frac14\Gamma^\rho_{\rho t} \n
& & \Gamma^r_{rr}=\frac38\frac{A^\prime}{A}-\frac18\frac{B^\prime}{B}
-\frac1{2r}+\frac14\Gamma^\rho_{\rho r},\;\;\Gamma^r_{r\theta}=
-\frac14\cot\theta+\frac14\Gamma^\rho_{\rho\theta},\;\;\Gamma^r_{r\phi}=
\frac14\Gamma^\rho_{\rho\phi}, \n
& & \Gamma^r_{\theta\theta}=-\frac{r}{2A}+\frac18\frac{r^2A^\prime}{A^2}
+\frac18\frac{r^2B^\prime}{AB}-\frac14\frac{r^2}{A}\Gamma^\rho_{\rho r},
\;\;\Gamma^r_{\phi\phi}=\sin^2\theta\Gamma^r_{\theta\theta}, \n
& & \Gamma^\theta_{tt}=-\frac B{4r^2}(\cot\theta-\Gamma^\rho_{\rho\theta})
,\;\;\Gamma^\theta_{t\theta}=\frac14\Gamma^\rho_{\rho t},\;\;
\Gamma^\theta_{rr}=\frac A{4r^2}(\cot\theta-\Gamma^\rho_{\rho\theta}),
\n
& & \Gamma^\theta_{r\theta}=\frac1{2r}-\frac{A^\prime}{8A}-\frac{B^\prime}
{8B}+\frac14\Gamma^\rho_{\rho r},\;\;\Gamma^\theta_{\theta\theta}=
-\frac14\cot\theta+\frac14\Gamma^\rho_{\rho\theta},\;\;
\Gamma^\theta_{\theta\phi}=\frac14\Gamma^\rho_{\rho\phi},\n
& & \Gamma^\theta_{\phi\phi}=-\frac34\sin\theta\cos\theta-\frac14
sin^2\theta\Gamma^\rho_{\rho\theta},\;\;\Gamma^\phi_{tt}=\frac14\frac
B{r^2\sin^2\theta}\Gamma^\rho_{\rho\phi},\;\;\Gamma^\phi_{t\phi}=
\frac14\Gamma^\rho_{\rho t},\n
& & \Gamma^\phi_{rr}=-\frac14\frac A{r^2\sin^2\theta}\Gamma^\rho_{\rho\phi},
\;\;\Gamma^\phi_{r\phi}=\frac1{2r}-\frac{A^\prime}{8A}-\frac{B^\prime}
{8B}+\frac14\Gamma^\rho_{\rho r},\;\;\Gamma^\phi_{\theta\theta}=
-\frac1{4\sin^2\theta}\Gamma^\rho_{\rho\phi},\n
& & \Gamma^\phi_{\phi\theta}=\frac34\cot\theta+\frac14\Gamma^\rho_
{\rho\theta},\;\;\Gamma^\phi_{\phi\phi}=\frac14\Gamma^\rho_{\rho\phi}.
\eeqa
($\prime$)denotes derivative with respect to r.\\By inserting Eq.(\ref{eq:24}) 
in Eq.(\ref{eq:12}) we obtain the components of Ricci tensor as follows:
\beqa 
& \R_{tt}= &-(\frac38\frac{B^\prime}{A}-\frac{BA^\prime}{8A^2}
-\frac B{2rA}+\frac B{4A}\Gamma^\rho_{\rho r})^\prime-\frac B
{4r^2}(1+\cot^2\theta+\frac\del{\del\theta}\Gamma^\rho_{\rho\theta})\n
& & +2(\frac38\frac{B^\prime}{B}-\frac{A^\prime}{8A}-\frac1{2r}+\frac14
\Gamma^\rho_{\rho r})(\frac38\frac{B^\prime}A-\frac{BA^\prime}{8A^2}
-\frac B{2rA}+\frac14\frac BA\Gamma^\rho_{\rho r})\n
& & +\frac B{8r^2}(-\cot\theta+\Gamma^\rho_{\rho\theta})^2-\frac18
\frac B{r^2\sin\theta}\Gamma^\rho_{\rho\phi}\n
& & -(\frac38\frac{B^\prime}A-\frac{BA^\prime}{8A^2}-\frac
B{2rA}+\frac B{4A}\Gamma^\rho_{\rho r})\Gamma^\rho_{\rho r}
-\frac B{4r^2}(-\cot\theta+\Gamma^\rho_{\rho\theta})\Gamma^\rho_
{\rho\theta} \label{eq:25}\\
& \R_{rr}= &-(\frac38\frac{A^\prime}A-\frac{B^\prime}{8B}-\frac1{2r}-
\frac34\Gamma^\rho_{\rho r})^\prime+\frac A{4r^2}(1+\cot^2\theta
+\frac\del{\del\theta}\Gamma^\rho_{\rho\theta})\n
& & +(\frac38\frac{B^\prime}B-\frac{A^\prime}{8A}-\frac1{2r}
+\frac14\Gamma^\rho_{\rho r})^2+\frac A{8B}{\Gamma^\rho_{\rho t}}^{^2}
+(\frac38\frac{A^\prime}A-\frac{B^\prime}{8B}-\frac1{2r}
+\frac14\Gamma^\rho_{\rho r})^2 \n
& & -\frac A{8r^2}(\cot\theta-\Gamma^\rho_{\rho\theta})^2+\frac
A{8r^2\sin^2\theta}{\Gamma^\rho_{\rho\phi}}^{^2}+2(\frac1{2r}
-\frac{A^\prime}{8A}-\frac{B^\prime}{8B}+\frac14\Gamma^\rho_{\rho r})^2
\n
& & -(\frac38\frac{A^\prime}A-\frac{B^\prime}{8B}-\frac1{2r}+\frac14
\Gamma^\rho_{\rho r})\Gamma^\rho_{\rho r}-\frac A{4r^2}(\cot\theta
-\Gamma^\rho_{\rho\theta})\Gamma^\rho_{\rho\theta} \label{eq:26}\\
& \R_{\theta\theta}= &\frac{\del\Gamma^\rho_{\rho\theta}}{\del\theta}
-(-\frac r{2A}+\frac{r^2A^\prime}{8A^2}+\frac{r^2B^\prime}{8AB}
-\frac{r^2}{4A}\Gamma^\rho_{\rho r})^\prime-\frac14(1+\cot^2\theta
+\frac{\del\Gamma^\rho_{\rho\theta}}{\del\theta})\n
& & \frac3{16}(-\cot\theta+\Gamma^\rho_{\rho\theta})^2-\frac
{r^2}{8B}{\Gamma^\rho_{\rho t}}^{^2}+\frac1{8\sin^2\theta}{\Gamma^
\rho_{\rho\phi}}^{^2}+(\frac34\cot\theta+\frac14\Gamma^\rho_{\rho\theta})^2\n
& & +2(-\frac r{2A}+\frac{r^2A^\prime}{8A^2}+\frac{r^2B^\prime}
{8AB}-\frac{r^2}{4A}\Gamma^\rho_{\rho r})(\frac1{2r}-\frac{A^\prime}
{8A}-\frac{B^\prime}{8B}+\frac14\Gamma^\rho_{\rho r})\n
& & -(-\frac r{2A}+\frac{r^2A^\prime}{8A^2}+\frac{r^2B^\prime}{8AB}
-\frac{r^2}{4A}\Gamma^\rho_{\rho r})\Gamma^\rho_{\rho r}+\frac14
(\cot\theta-\Gamma^\rho_{\rho\theta})\Gamma^\rho_{\rho\theta} \label{eq:27}\\
& \R_{\phi\phi}= &-\sin^2\theta(-\frac r{2A}+\frac{r^2A^\prime}{8A^2}+\frac{r^2B^\prime}
{8AB}-\frac{r^2}{4A}\Gamma^\rho_{\rho r})^\prime\n
& & +\frac34\cos 2\theta+\frac14\frac\del{\del\theta}(\sin^2\theta
\Gamma^\rho_{\rho\theta})-\frac{r^2\sin^2\theta}{8B}{\Gamma^\rho_
{\rho t}}^{^2}\n
& & +2\sin^2\theta(-\frac{r}{2A}+\frac{r^2A^\prime}{8A^2}+\frac{r^2B^\prime}{8AB}
-\frac{r^2}{4A}\Gamma^\rho_{\rho r})
(\frac{1}{2r}-\frac{A^\prime}{8A}-\frac{B^\prime}{8B}+\frac 14\Gamma^\rho_{\rho r})\n
& &-2\sin^2\theta(\frac34\cot\theta+\frac14\Gamma^\rho_{\rho\theta})^2
-\sin^2\theta(-\frac r{2A}+\frac{r^2A^\prime}{8A^2}+\frac{r^2B^\prime}
{8AB}-\frac{r^2}{4A}\Gamma^\rho_{\rho r})\Gamma^\rho_{\rho r}\n
& &+\sin^2\theta(\frac34\cot\theta+\frac14\Gamma^\rho_{\rho\theta})
\Gamma^\rho_{\rho\theta} \label{eq:28}.
\eeqa
As it is evident from Eqs.(\ref{eq:25})-(\ref{eq:28}) these are functions of $\theta$
which is not consistent with the symmetry of spacetime. But if we choose
$\Gamma^\rho_{\rho t}=\Gamma^\rho_{\rho\phi}=0$ , $\Gamma^\rho_{\rho
\theta}=\cot\theta$ and $\Gamma^\rho_{\rho r}=\Gamma^\rho_{\rho r}(r)$
then the Ricci tensor shows its symmetry manifestly. Imposing these values for
the unspecified components of trace of the connection in Eqs.(\ref{eq:25})-(\ref{eq:28})
leads to the following results:
\beqa
& \R_{tt}= &-(\frac38\frac{B^\prime}A-\frac{BA^\prime}{8A^2}-\frac{B}{2rA}
+\frac{B}{4A}\Gamma^\rho_{\rho r})^\prime\n
& & +2(\frac{3B^\prime}{8B}-\frac{A^\prime}{8A}-\frac1{2r}+\frac14\Gamma^\rho_{\rho r})
(\frac{3B^\prime}{8A}-\frac{BA^\prime}{8A^2}-\frac B{2rA}+\frac B{4A}\Gamma^\rho_{\rho r})
\n
& & -\Gamma^\rho_{\rho r}(\frac38\frac{B^\prime}{8A}-\frac18\frac{BA^\prime}{A^2}-\frac B{2rA}+\frac14\Gamma^\rho_{\rho r}\frac BA) \label{eq:29}\\
& \R_{rr}= &-(\frac{3A^\prime}{8A}-\frac{B^\prime}{8B}-\frac1{2r}-\frac34\Gamma^\rho_
{\rho r})^\prime+(\frac{3B^\prime}{8B}-\frac{A^\prime}{8A}-\frac1{2r}+\frac14\Gamma^\rho_
{\rho r})^2 \n
& & +(\frac{3A^\prime}{8A}-\frac{B^\prime}{8B}-\frac1{2r}+\frac14\Gamma^\rho_{\rho r})^2
+2(\frac1{2r}-\frac{A^\prime}{8A}-\frac{B^\prime}{8B}+\frac14\Gamma^\rho_{\rho r})^2
\n
& & -(\frac{3A^\prime}{8A}-\frac{B^\prime}{8B}-\frac1{2r}+\frac14\Gamma^\rho_{\rho r})
\Gamma^\rho_{\rho r} \label{eq:30}\\
& \R_{\theta\theta}= &-1-(-\frac r{2A}+\frac{r^2A^\prime}{8A^2}+\frac{r^2B^\prime}
{8AB}-\frac{r^2}{4A}\Gamma^\rho_{\rho r})^\prime\n
& & +2(-\frac r{2A}+\frac{r^2A^\prime}{8A^2}+\frac{r^2B^\prime}{8AB}-\frac{r^2}{4A}
\Gamma^\rho_{\rho r})(\frac1{2r}-\frac{A^\prime}{8A}-\frac{B^\prime}{8B}+\frac14\Gamma^
\rho_{\rho r})\n
& & -(-\frac r{2A}+\frac{r^2A^\prime}{8A^2}+\frac{r^2B^\prime}{8AB}-\frac{r^2}{4A}\Gamma^\rho
_{\rho r}) \label{eq:31}\\
& \R_{\phi\phi}= &\sin^2\theta \;\R_{\theta\theta} \label{eq:32}
\eeqa 
By using Eqs.(\ref{eq:30})-(\ref{eq:32}),(\ref{eq:13})and(\ref{eq:23}) with some manipulation \R $\;$ takes the form;
\beqa
& \R= &\frac{B^\frac14 r\sin^\frac12\theta}{A^\frac34}\{\frac32{\Gamma^\rho_{\rho r}}^{^\prime}
+\frac{B^{\prime\prime}}{4B}-\frac{3A^{\prime\prime}}{4A}+(\frac12-2A)\frac1{r^2}
+\frac{39}{32}(\frac{A^\prime}{A})^2\n
& & -\frac1{32}(\frac{B^\prime}B)^2-\frac5{16}\frac{A^\prime B^\prime}{AB}-\frac
{5A^\prime}{4rA}-\frac{B^\prime}{4rB}+\frac3{2r}\Gamma^\rho_{\rho r}\n
& & -\frac{9A^\prime}{8A}\Gamma^\rho_{\rho r}+\frac{3B^\prime}{8B}\Gamma^\rho_{\rho r}
+\frac38{\Gamma^\rho_{\rho r}}^{^2}\} \label{eq:33}
\eeqa 
Multiplying the $tt$ component of the field equation (\ref{eq:20}) by $\frac{4}{B}$ and substituting Eqs.(\ref{eq:29})-(\ref{eq:31}) in it leads to
\beqa
& & \frac12{\Gamma^\rho_{\rho r}}^{^\prime}-\frac{5B^{\prime\prime}}{4B}
-\frac{A^{\prime\prime}}{4A}+(\frac12-2A)\frac1{r^2}+\frac{11}{32}(\frac{A^\prime}{A})^2
+\frac{35}{32}(\frac{B^\prime}{B})^2+\frac{15}{16}\frac{A^\prime B^\prime}{AB}\n
& & -\frac9{4r}\frac{A^\prime}A
-\frac5{4r}\frac{B^\prime}B+\frac3{2r}\Gamma^\rho_{\rho r}-\frac{A^\prime}{8A}
\Gamma^\rho_{\rho r}-\frac{5B^\prime}{8B}\Gamma^\rho_{\rho r}
-\frac18{\Gamma^\rho_{\rho r}}^{^2}=0 \label{eq:34}
\eeqa
Multiplying the $rr$ component of Eq.(\ref{eq:20}) by $\frac{4}{A}$ and substituting Eqs.(\ref{eq:29})-(\ref{eq:31})in it gives
\beqa
& & \frac32{\Gamma^\rho_{\rho r}}^{^\prime}+\frac{B^{\prime\prime}}{4B}-\frac
{3A^{\prime\prime}}{4A}+\frac3{2r^2}+\frac{33}{32}(\frac{A^\prime}A)^2
+\frac9{32}(\frac{B^\prime}B)^2-\frac{3A^\prime B^\prime}{16AB}\n
& & -\frac{3A^\prime}{4rA}-\frac{7B^\prime}{4rB}
+\frac1{2r}\Gamma^\rho_{\rho r} \label{eq:35}
+\frac{B^\prime}{8B}\Gamma^\rho_{\rho r}
-\frac38\frac{A^\prime}A\Gamma^\rho_{\rho r}+\frac{2A}{r^2}
-\frac38{\Gamma^\rho_{\rho r}}^{^2}=0
\eeqa
Multiplication of the $\theta\theta$ component of Eq.(\ref{eq:20}) by $\frac{4}{r^2}$ and insertion of Eqs.(\ref{eq:29})-(\ref{eq:31}) in it gives the final relation
\beqa
& & -\frac12{\Gamma^\rho_{\rho r}}^{^\prime}-\frac{3B^{\prime\prime}}{4B}+
\frac{A^{\prime\prime}}{4A}-(\frac12+2A)\frac1{r^2}-\frac{11}{32}(\frac{A^\prime}A)^2
+\frac{13}{32}(\frac{B^\prime}B)^2+\frac{9A^\prime B^\prime}{16AB}\n
& & -\frac{3A^\prime}{4rA}+\frac{B^\prime}{4rB}+\frac1{2r}\Gamma^\rho_{\rho r}
+\frac{A^\prime}{8A}\Gamma^\rho_{\rho r}-\frac{3B^\prime}{8B}\Gamma^\rho_{\rho r}
+\frac18{\Gamma^\rho_{\rho r}}^{^2}=0 \label{eq:36}
\eeqa
The $\phi\phi$ component of Eq.(\ref{eq:20}) does not yield to a new equation and Eq.(\ref{eq:36}) 
is repeated. It is interesting to notice that Eqs.(\ref{eq:34}),(\ref{eq:35}) and (\ref{eq:36})
are not independent and we have indeed two independent relations.
If we add Eq.(\ref{eq:34}) and Eq.(\ref{eq:36}) will result in
\beq \label{eq:37}
\frac{B^{\prime\prime}}B=-\frac{2A}{r^2}+\frac34(\frac{B^\prime}B)^2+
\frac{3A^\prime B^\prime}{4AB}-\frac{3A^\prime}{2rA}-\frac{B^\prime}{2rB}
-\frac{B^\prime}{2B}\Gamma^\rho_{\rho r}+\frac1r\Gamma^\rho_{\rho r}
\eeq
Also addition of Eq.(\ref{eq:34}) and (\ref{eq:35}) and substitution of the value of
$\frac{B^{\prime\prime}}B$ from Eq.(\ref{eq:37}) gives
\beqa
\frac{A^{\prime\prime}}A & = & 2{\Gamma^\rho_{\rho r}}^{^\prime}+
\frac{2(1+A)}{r^2}+\frac{11}8(\frac{A^\prime}A)^2+\frac58(\frac{B^\prime}B)^2
-\frac{3A^\prime}{2rA}\n
& & -\frac{5B^\prime}{2rB}+\frac1r\Gamma^\rho_{\rho r}-\frac{A^\prime}{2A}
\Gamma^\rho_{\rho r}-\frac12{\Gamma^\rho_{\rho r}}^{^2} \label{eq:38}
\eeqa
Using Eq.(\ref{eq:35}), Eq.(\ref{eq:33}) for \R $\;$can be written as
\beqa
& \R= &\frac{B^\frac14 r \sin^\frac12\theta}{A^\frac34}\lbrack -(1+4A)\frac1{r^2}+\frac
3{16}(\frac{A^\prime}A)^2-\frac5{16}(\frac{B^\prime}B)^2-\frac{A^\prime B^\prime}
{8AB}-\frac{A^\prime}{2rA}\n
& & +\frac{3B^\prime}{2rB}+\frac1r\Gamma^\rho_{\rho r}-\frac{3A^\prime}{4A}\Gamma
^\rho_{\rho r}+\frac{B^\prime}{4B}\Gamma^\rho_{\rho r}+\frac34{\Gamma^\rho_{\rho r}}^
{^2}\rbrack. \label{eq:39}
\eeqa
Now field equation (\ref{eq:19}) will be satisfied automatically if we insert
Eq.(\ref{eq:39}) in Eq.(\ref{eq:19}) and replace $\frac{B^{\prime\prime}}B$ and $\frac{A^
{\prime\prime}}A$ from Eqs.(\ref{eq:37}) and (\ref{eq:38}) in it. This means that the 
field equation (\ref{eq:19}) does not lead to an independent relation.\\ 
\indent An empty space is spherically symmetric at each point so it is 
homogeneous and isotropic. For an empty space $A$ abd $B$ are independent
of radial coordinate and consequently are constant. Then Eqs.(\ref{eq:37},\ref{eq:39})
with constant $A$ and $B$ imply $\Gamma^\rho_{\rho r}=\frac 2r$ for this case.
The action for a vacuum spherical symmetry space is the same as
the action of an empty space plus an additional matter term corresponding
to a single point source which its existence at the origin is the 
cause of spherical symmetry. Actually this additional term
does not depend on $\Gamma^\rho_{\rho \lambda}$. So the field 
equations corresponding to the variation with respect to
$\Gamma^\rho_{\rho\lambda}$ are the same for the both cases.
Meanwhile both these cases satisfy the same boundary conditions.
So this necessitates the same result for $\Gamma^\rho_{\rho\lambda}$
in both cases. A rigorous proof for $\Gamma^\rho_{\rho r}=\frac 2r$ will be given in the next section in the context of a gauge invariance discussion.

\section{Gauge Fixing}
First we demonstrate that at any point P there exists a locally inertial coordinate $x^\prime$ in which $\g_{\mu ^\prime\nu ^\prime}$ takes its canonical form $\eta_{\mu ^\prime\nu ^\prime}$ and the first partial derivatives of the metric $\del_{\sigma ^\prime}\g_{\mu ^\prime\nu ^\prime}$ and the components of the trace of the connection $\Gamma ^{\rho ^\prime}_{\rho ^{\prime}\lambda ^{\prime}}$ all vanish. Meanwhile some of the components of the second partial derivatives of the metric $\del_{\rho ^\prime}\del_{\sigma ^\prime}\g_{\mu ^\prime\nu ^\prime}$ and the first partial derivative of the trace of the connection can not be made to vanish. Let us consider the transformation law for the metric and the trace of the connection;
\beqa 
\g_{\mu ^\prime\nu ^\prime}& = &|\frac{\del x}{\del x^\prime}|^{^{-\frac12}}\frac{\del x^\mu}{\del x^{\mu ^\prime}}\frac{\del x^\nu}{\del x^{\nu ^\prime}}\g_{\mu\nu} \label{eq.46-1}\\
\Gamma ^{\rho ^\prime}_{\rho ^{\prime}\lambda ^{\prime}}& = &\frac{\del x^\lambda}{\del x^{\lambda ^\prime}}\Gamma ^\rho_{\rho\lambda}+\frac{\del x^{\nu ^\prime}}{\del x^\lambda}\frac{\del ^2 x^\lambda}{\del x^{\nu ^\prime}\del x^{\lambda ^\prime}} \label{eq.46-2}
\eeqa
and expand both sides in Taylor series in the sought-after coordinate $x^{\mu ^\prime}$. The expansion of the old coordinate looks like
\beq \label{eq.46-3}
x^\mu=(\frac{\del x^\mu}{\del x^{\mu ^\prime}})_{_P} x^{\mu ^\prime}
+\frac12(\frac{\del^2 x^\mu}{\del x^{\mu ^{\prime}_1}\del x^{\mu ^{\prime}_2}})_{_P} x^{\mu ^{\prime}_1} x^{\mu ^{\prime}_2}+\frac16(\frac{\del ^3x^\mu}{\del x^{\mu ^{\prime}_1}\del x^{\mu ^{\prime}_2}\del x^{\mu ^{\prime}_3}})x^{\mu ^{\prime}_1}x^{\mu ^{\prime}_2}x^{\mu ^{\prime}_3}+ ...
\eeq 
(For simplicity we have $x^\mu (P)=x^{\mu ^{\prime}}(P) =0$).
Then using some schematic notation , the expansion of Eqs.(\ref{eq.46-1})and (\ref{eq.46-2}) to second order are
\beqa
& &(\g^\prime)_P+(\del ^\prime \g^\prime)_P x^\prime+(\del ^\prime\del ^\prime \g^\prime)_P x^\prime x^\prime+...\n
& &=(|\frac{\del x}{\del x^\prime}|^{-\frac12}\frac{\del x}{\del x^\prime}\frac{\del x}{\del x^\prime} \g)_P+[|\frac{\del x}{\del x^\prime}|^{-\frac12}(2\frac{\del x}{\del x^\prime}\frac{\del ^2x}{\del x^\prime\del x^\prime}\g+\frac{\del x}{\del x^\prime}\frac{\del x}{\del x^\prime}\del ^\prime \g \n 
& &-\frac12\frac{\del x^\prime}{\del x}\frac{\del ^2x}{\del x^\prime\del x^\prime}\frac{\del x}{\del x^\prime}\frac{\del x}{\del x^\prime}\g)]_P x^\prime+... \label{eq.46-4}
\eeqa 
\beqa
& &(\Gamma^\prime)_P+(\del ^\prime\Gamma^\prime)_P x^\prime+... \n
& &=[\frac{\del x}{\del x^\prime}\Gamma +\frac{\del x^\prime}{\del x}\frac{\del ^2 x}{\del x^\prime\del x^\prime}]_P+[\frac{\del ^2 x}{\del x^\prime\del x^\prime}\Gamma+\frac{\del x}{\del x^\prime}\del ^\prime\Gamma \n
& &+\frac{\del x}{\del x^\prime}\frac{\del ^3 x}{\del x^\prime\del x^\prime\del x^\prime}+\frac{\del ^2 x}{\del x^\prime\del x^\prime}\frac{\del ^2 x^\prime}{\del x \del x}\frac{\del x}{\del x^\prime}]_P x^\prime+... \label{eq.46-5}
\eeqa
We can set terms of the same order in $x^\prime$ on each side to be equal. Therefore according to Eq.(\ref{eq.46-4}),the components $g_{\mu ^\prime\nu ^\prime}$, ten numbers in all are determined by the matrix $(\frac{\del x^\mu}{\del x^{\mu ^\prime}})_P$. This is a $4\times4$ matrix with no constraint. Thus we are free to choose sixteen numbers. This is enough freedom to put the ten numbers of $g_{\mu ^\prime\nu ^\prime}(P)$ into canonical form $\eta_{\mu ^\prime\nu ^\prime}$.
The six remained degrees of freedom can be interpreted as exactly the six parameters of the Lorentz group that these leave the canonical form unchanged. At first order we have the derivative $\del_{\sigma^\prime}g_{\mu^\prime\nu^\prime}(P)$, four derivatives of ten components for a total of forty numbers. Since $\g^{\mu\nu}\del_\lambda \g_{\mu\nu}=0$ we have merely 36 independent numbers for them. Now looking at the right hand side of Eq.(\ref{eq.46-4}), we have the additional freedom to choose $\frac{\del^2 x}{\del x^{\mu^\prime_1}\del x^{\mu^\prime_2}}$.
In this set of numbers there is a total number of 40 degrees of freedom. Precisely the number of choices we need to determine all of the first derivatives of the metric is 36, which we can set to zero, while 4 numbers of the 40 degrees of freedom of $\frac{\del^2 x^\mu}{\del x^{\mu^\prime_1}\del x^{\mu^\prime_2}}$ remain unused. According to Eq.(\ref{eq.46-5}) the four components of $\Gamma^{\rho^\prime}_{\rho^\prime\lambda^\prime}(P)$ may be determined by the four remaining components of the matrix $(\frac{\del^2 x^\mu}{\del x^{\mu^\prime_1}\del x^{\mu^\prime_2}})_P $ which we can set them equal to zero too. We should emphasized that $\Gamma^\rho_{\rho\lambda}=0$ in polar coordinates takes the form of $(0,\frac2r.\cot\theta,0)$. So it is always possible to find a locally inertial coordinate system in which the components of the Riemann curvature tensor take the form;
\beqa
\R_{\mu\nu\rho\sigma} & {\buildrel\star\over=}& \eta_{\mu\lambda}(\del_\rho\Gamma^\lambda_{\nu\sigma}-\del_\sigma\Gamma^\lambda_{\nu\rho}) \n
&{\buildrel\star\over=}& \frac12 (\del_\rho\del_\nu \g_{\mu\sigma}+\del_\sigma\del_\mu \g_{\nu\rho}-\del_\sigma\del_\nu \g_{\mu\rho}-\del_\rho\del_\mu \g_{\nu\sigma}) \n
& & +\frac14 (\eta_{\mu\sigma}\del_\rho\Gamma^\xi_{\xi\nu}+\eta_{\mu\nu}\del_\rho\Gamma^\xi_{\xi\sigma}-\eta_{\nu\sigma}\del_\rho\Gamma^\xi_{\xi\mu} \n
& &-\eta_{\mu\rho}\del_\sigma\Gamma^\xi_{\xi\nu}-\eta_{\mu\nu}\del_\sigma\Gamma^\xi_{\xi\rho}+\eta_{\nu\rho}\del_\sigma\Gamma^\xi_{\xi\mu})\label{eq.46-6}
\eeqa 
The Ricci tensor comes from the contracting over $\mu$ and $\rho$ , 
\beq \label{eq.46-7}
\R_{\nu\sigma}\;{\buildrel\star\over=} \; \eta^{\mu\rho}\R_{\mu\nu\rho\sigma}
\eeq
The change of the metric which is a density tensor of rank $(0,2)$ and weight $-\frac 12$ under a diffeomorphism along the vector field $\xi^\mu$ is 
\beq \label{eq.46-8}
\delta\g_{\mu\nu}={\cal{L}}_\xi\g_{\mu\nu}=
\nabla_\mu\xi_\nu+\nabla_\nu\xi_\mu-\frac12\g_{\mu\nu}
\nabla_\lambda\xi^\lambda
\eeq
Eq.(\ref{eq.46-8}) in the local inertial frame takes the form
\beq\label{eq.46-9}
\delta\g_{\mu\nu}\; {\buildrel\star\over=}\; \del_\mu\xi_\nu+\del_\nu\xi_\mu -\frac12\eta_{\mu\nu}\del_\lambda\xi^\lambda
\eeq
The change of the trace of the connection in local inertial frame under this diffeomorphism is
\beq \label{eq.46-10}
\delta\Gamma^\rho_{\rho\mu}\; {\buildrel\star\over=}\; \del_\mu\del_\lambda\xi^\lambda
\eeq
We may call Eqs.(\ref{eq.46-9}) and(\ref{eq.46-10}) a gauge transformation.
Ricci tensor Eq.(\ref{eq.46-7}) under the gauge transformation (\ref{eq.46-9}),
(\ref{eq.46-10}) is invariant. This gauge degree of freedom has to be fixed before going further. This may be achieved by taking 
\beq \label{eq.46-11}
\Gamma^\rho_{\rho\mu}=0
\eeq
As it has been mentioned the gauge (\ref{eq.46-11}) in polar coordinates leads to
\beq \label{eq.46-12}
\Gamma^\rho_{\rho\mu}=(0,\frac2r,\cot\theta,0)
\eeq
So we take $\Gamma^\rho_{\rho r}=\frac2r$ without any lose of generality in our calculations. This spacetime has four killing vectors just the same as Schwarzschild spacetime, one timelike and three spacelike that are the same as those on $S^2$.

\section{$\Gamma^\rho_{\rho r}=\frac 2r$}
Now let us investigate
the solutions of Eqs.(\ref{eq:37}) and (\ref{eq:38}) for the given form of $\Gamma^\rho
_{\rho r}=\frac2r$.  This leads to the following two relations for
$A$ and $B$:
\beqa
\frac{B^{\prime\prime}}B& = &\frac{2(1-A)}{r^2}+\frac34(\frac{B^\prime}B)^2
+\frac34\frac{A^\prime B^\prime}{AB}-\frac{3A^\prime}{2rA}-\frac{3B^\prime}{2rB} \label{eq:41}\\
\frac{A^{\prime\prime}}A& = &\frac{2(-1+A)}{r^2}+\frac{11}8(\frac{A^\prime}A)^2
+\frac58(\frac{B^\prime}B)^2-\frac{5A^\prime}{2rA}-\frac{5B^\prime}{2rB} \label{eq:42}
\eeqa 
Combining Eqs.(\ref{eq:41})and (\ref{eq:42})with some manipulation gives
\beq \label{eq:43}
(\frac{A^\prime}A+\frac{B^\prime}B)^{^\prime}=\frac38(\frac{A^\prime}A+\frac{B^\prime}B)
^{^2}-\frac4r(\frac{A^\prime}A+\frac{B^\prime}B)
\eeq
Eq.(\ref{eq:43})can be easily integrated with respect to $r$, then it results
\beq \label{eq:44}
(\frac{A^\prime}A+\frac{B^\prime}B)=
\frac{C(AB)^\frac38}{r^4}
\eeq
where $C$ is constant of integration. Now if we introduce the new parameter
$y=AB$, then integration of Eq.(\ref{eq:44}) with respect to $r$ gives the solution
\beq \label{eq:45}
y^{-\frac38}=\frac{C}{8r^3}+\hat{C}
\eeq
where $\hat{C}$ is another integration constant. To be consistent with the asymptotic form of $A=B=1$ at infinity we should have $\hat{C}=1$
\beq \label{eq:46}
AB=(1+\frac{C}{8r^3})^{-\frac83}
\eeq
Then by inserting Eq.(\ref{eq:46}) into Eq.(\ref{eq:41}) we find an equation for $B$,
\beq \label{eq:47}
B^{\prime\prime}=\frac{2[B-r^8(r^3+\frac C8)^{-\frac83}]}{r^2}
+\frac{C(\frac34 rB^\prime-\frac32 B)}{r^2(r^3+\frac C8)}
\eeq
Eq.(\ref{eq:47}) has a general solution as follows:
\beq \label{eq:48}
B(r)=\frac1{(1+\frac{C^\prime}{r^3})^\frac23}+\frac{\alpha r^2}{1+\frac{C^\prime}{r^3}}
+\frac{\beta}{(1+\frac{C^\prime}{r^3})r}
\eeq
where $C^\prime=\frac C8$ , $\alpha$ and$\beta$ are constants of integration.\\
Newtonian limit requires that,($c=1$)
\beq \label{eq:49}
B\;\rightarrow\;1-\frac{2GM}r \hspace{1cm} \mbox{as}\;\; r\;\rightarrow\;\infty
\eeq  
Applying condition (\ref{eq:49}) to the Eq.(\ref{eq:48}) implies that
\beq \label{eq:50}
\beta-\alpha C^\prime= -2GM
\eeq
Here $\alpha$ has the dimension $[L]^{-2}$ and may be called the 
cosmological constant. So we may take
\beq \label{eq:51}
\beta=-2GM+\Lambda C^\prime
\eeq
and we will have
\beqa
& B(r)= &(1+\frac{C^\prime}{r^3})^{-\frac23}[1-\frac{2GM-\Lambda C^\prime}r
(1+\frac{C^\prime}{r^3})^{-\frac13}+\Lambda r^2(1+\frac{C^\prime}{r^3})^{-\frac13}] \label{eq:52}\\
& A(r)= &(1+\frac{C^\prime}{r^3})^{-2}[1-\frac{2GM-\Lambda C^\prime}r
(1+\frac{C^\prime}{r^3})^{-\frac13}+\Lambda r^2(1+\frac{C^\prime}{r^3})^{-\frac13}]^{-1} \label{eq:53}
\eeqa
A special case of this general solution is when $\Lambda=0$,
\beqa
& B(r)= &(1+\frac{C^\prime}{r^3})^{-\frac23}[1-\frac{2GM}r(1+\frac{C^\prime}{r^3})^
{-\frac13}] \label{eq:54}\\
& A(r)= &(1+\frac{C^\prime}{r^3})^{-2}[1-\frac{2GM}r(1+\frac{C^\prime}{r^3})^{-\frac13}]^{-1} \label{eq:55}
\eeqa
Eqs.(\ref{eq:54}) and (\ref{eq:55}) show that the case $C^\prime=0$ gives the familiar
Schwarzschild solution.\\
The term $(1+\frac{C^\prime}{r^3})$ is always nonnegative for $C^\prime\geq 0$, so $B(r)$ and $A(r)$
are always nonnegative if the condition
\beq \label{eq:56}
1-\frac{2GM}r(1+\frac{C^\prime}{r^3})^{-\frac13}\geq 0
\eeq
holds for the hole range of $r$. Eq.(\ref{eq:56}) can be rewritten as
\beq \label{eq:57}
r^3\geq (2GM)^3-C^\prime
\eeq
Eq.(\ref{eq:57}) implies that Eq.(\ref{eq:56}) will be true for the hole range of $r$
if we take
\beq \label{eq:58}
C^\prime \geq (2GM)^3
\eeq
If we assume that the condition (\ref{eq:58}) holds then $A$ and $B$
will be analytic on the hole range of $r$. A remarkable point about $\Lambda$
which has the same role of cosmological constant is that choosing $\Lambda=0$
is merely to simplify the calculations. Keeping $\Lambda \ne 0$ does not change the character of $C^\prime$ significantly. We omit the detail of calculations to avoid mathematical complications. The solution asymptotically becomes deSitter-Schwarzschild.\\ 
\indent Considering a null radial geodesic, its line element gives
\beq \label{eq:59}
\pm dt=\sqrt{\frac AB}dr
\eeq
Putting Eq.(\ref{eq:54})and Eq.(\ref{eq:55})in Eq.(\ref{eq:59}) gives
\beq
\pm dt=\frac{r^2dr}
{(r^3+C^\prime)^{\frac13}((r^3+C^\prime)^{\frac13}-2GM)} \label{eq:60}
\eeq
Let us define the new parameter $R=(r^3+C^\prime)^\frac13$.
The range of $R$ will be from $(C^\prime)^\frac13$ to infinity.
We have
\beq \label{eq:61}
R^2dR=r^2dr
\eeq
Then Eq.(\ref{eq:60}) takes the form
\beq \label{eq:62}
\pm dt=\frac{RdR}{R-2GM}
\eeq
Integrating Eq.(\ref{eq:62}) gives
\beqa
& \pm(t-t_0)= &(r^3+C^\prime)^\frac13+2GM\ln[(r^3+C^\prime)^\frac13-2GM]\n
& &-(r_0^3+C^\prime)^\frac13+2GM\ln[(r_0^3+C^\prime)^\frac13-2GM] \label{eq:63}
\eeqa
Since we have assumed that $C^\prime>(2GM)^3$ , then Eq.(\ref{eq:63}) shows no sign of singularity on the hole range of $r$.

\section{Upper bound on $C^\prime$}
It is natural to anticipate the classical test like the advance of the perihelion of Mercury to put some upper bound on the probable values of $C^\prime$. Fortunately high degree of symmetry greatly simplify our task. There are four Killing vectors which each of these will lead to a constant of the motion for a free particle. In addition we always have another constant of the motion for the geodesics; the geodesic equation implies that the quantity 
\beq \label{eq:64}
\epsilon = -g_{\mu\nu}\frac{dx^\mu}{d\lambda}\frac{dx^\nu}{d\lambda}
\eeq
is constant along the path. For massive particles we typically choose $\lambda$ so that $\epsilon=1$. For massless particles we always have $\epsilon=0$. Invariance under time translations leads to conservation of energy while invariance under spatial rotations leads to conservation of the three components of angular momentum. Conservation of direction of angular momentum means that particle moves on a plane. So we may choose $\theta =\frac{\pi}{2}$. The two remaining Killing vectors correspond to the energy and the magnitude of angular momentum. The energy arises from the timelike Killing vector
\beq
(\partial_t)^\mu = (1,0,0,0). \label{eq:65}
\eeq  
While the Killing vector whose conserved quantity is the magnitude of angular momentum is
\beq \label{eq:66}
(\partial_\phi)^\mu = (0,0,0,1)
\eeq 
The two conserved quantities are :
\beq
E=-(1+\frac{C^\prime}{r^3})^{-\frac23}[1-\frac{2GM}{r}(1+\frac{C^\prime}{r^3})^{-\frac13}]\frac{dt}{d\lambda} \label{eq:67}
\eeq 
and
\beq
L=r^2\frac{d\phi}{d\lambda}\label{eq:68}
\eeq  
Expanding the expression(\ref{eq:64}) and multiplying it by Eq.(\ref{eq:54})
and using Eq.(\ref{eq:67}) and Eq.(\ref{eq:68})give 
\beq 
-E^2+(1+\frac{C^\prime}{r^3})^{-\frac83}(\frac{dr}{d\lambda})^2+(1+\frac{C^\prime}{r^3})^{-\frac23}(1-\frac{2GM}{r}(1+\frac{C^\prime}{r^3})^{-\frac13})(\frac{L^2}{r^2}+\epsilon)=0 \label{eq:69}
\eeq    
Expanding in powers of $r$ , we shall content ourselves here with only the lowest order of approximation in Eq.(\ref{eq:69}). Then we have
\beq \label{eq:70}
(\frac{dr}{d\lambda})^2+(1-\frac{2GM}{r})(\frac{L^2}{r^2}+\epsilon)
-E^2(1+\frac83\frac{C^\prime}{r^3})+\frac{2\epsilon C^\prime}{r^3}=0
\eeq 
Now by taking $\epsilon=1$ and multiplying Eq.(\ref{eq:70}) by $(\frac{d\phi}{d\lambda})^{-2}=\frac{r^4}{L^2}$ and defining 
$x=\frac{L^2}{GMr}$ we have:
\beq
(\frac{dx}{d\phi})^2-2x+x^2-\frac{2G^2M^2}{L^2}x^3+\frac{GM}{L^4}(2C^\prime-\frac{8E^2C^\prime}{3})x^3=\frac{E^2L^2}{G^2M^2}-\frac{L^2}{G^2M^2}
\label{eq:71}
\eeq
Differentiating Eq.(\ref{eq:71}) with respect to $\phi$ gives
\beq
\frac{d^2x}{d\phi^2}-1+x=(\frac{3G^2M^2}{L^2}-\frac{3GM}{L^4}C^\prime+\frac{4E^2C^\prime GM}{L^4})x^2 \label{eq:72}
\eeq 
After some manipulation we obtain that the perihelion advances by an angle
\beq
\Delta\phi=\frac{6\pi G^2M^2}{L^2}(1-\frac{C^\prime}{GML^2}+\frac{4C^\prime E^2}{3GML^2}) \label{eq:73}
\eeq 
On the other hand we have
\beq
E^2=1+\frac{G^2M^2}{L^2}(e^2-1) \label{eq:74}
\eeq 
and
\beq
L^2\cong GM(1-e^2)a \label{eq:75}
\eeq 
where $e$ is the eccentricity and $a$ is the semi-major axis of ellipse.
Putting Eqs.(\ref{eq:74}) and (\ref{eq:75}) into Eq.(\ref{eq:73}) yields
\beq
\Delta\phi=\frac{6\pi GM}{(1-e^2)a}(1+\frac{C^\prime}{3G^2M^2(1-e^2)a}+\frac{4C^\prime}{3GM(1-e^2)a^2})
\label{eq:76}
\eeq 
The third term in the parenthesis is smaller than the second term by a $\frac{GM}{a}$
factor and may be neglected.
\beq
\Delta\phi=\frac{6\pi GM}{(1-e^2)a}(1+\frac{C^\prime}{3G^2M^2(1-e^2)a}) \label{eq:77}
\eeq
This shows a sever dependence on $C^\prime$ and should not conflict with observational data, i.e. we should have
\beq
C^\prime<G^2M^2a \label{eq:78}
\eeq 
by an order of magnitude. 
\section{Conclusions and Remarks}
A new field equation for gravity is proposed by introducing a new character for metric as a tensor density in the absence of matter. We obtained the vacuum spherical symmetry solutions of the field equations. A considerable large part of these solutions are regular for the whole range of $r$ except at $r=0$. Since
$\partial_t$ remains timelike everywhere, so there is no event horizon in the spacetime. The solutions for the metric show a singularity at $r=0$. To find out
the nature of the singularity we need to check some scalar densities made of 
Riemann tensor. First let us check the behavior of Riemann scalar density of weight
$+\frac 12$ , $\R$. Substituting $\Gamma^\rho_{\rho r}=\frac 2r$ in Eq.(\ref{eq:39})
results in
\beq \label{eq:79}
\R =\frac{B^{\frac 14}r\sin^{\frac 12}\theta}{A^{\frac 34}}\left\{\frac{4(1-A)}{r^2}
+(\frac{B^\prime}{B}-\frac{A^\prime}{A})[\frac 2r-\frac 14(\frac{A^\prime}{A}
+\frac{B^\prime}{B})]-\frac{1}{16}(\frac{A^\prime}{A}+\frac{B^\prime}{B})^2\right\}
\eeq 
In the range of Newtonian limit when $r>>2GM$ , $A$ and $B$ in Eqs.(\ref{eq:54}),(\ref{eq:55}) reduce to the Schwarzschild form, so Eq.(\ref{eq:79})
gives $\R=0$. For the range of $r<<2GM$ we have two different asymptotic forms for 
$A$ and $B$ depending on the value of $C^\prime$.
If $C^\prime>(2GM)^3$  then we have
\beqa
B&\approx& (C^\prime)^{-\frac23}(1-\frac{2GM}{(C^\prime)^{\frac13}})r^2 \label{eq:80}\\
A&\approx& (C^\prime)^{-2}(1-\frac{2GM}{(C^\prime)^\frac13})r^6 \label{eq:81}
\eeqa
For $C^\prime=(2GM)^3$ , $A$ and $B$ reduce to 
\beqa
B&=&\frac13(\frac{r}{2GM})^3 \label{eq:82}\\
A&=&3(\frac{r}{2GM})^5 \label{eq:83}
\eeqa
For both cases by inserting Eqs.(\ref{eq:80})-(\ref{eq:83}), into Eq.(\ref{eq:79})
,$\R$ becomes
\beq  \label{eq:84}
\R=-4(C^\prime)^{-\frac23}r\sin^\frac12\theta
\eeq 
Immediately we may calculate $\R^{\mu\nu}\R_{\mu\nu}$. According to the field equation 
(\ref{eq:20}) we have $\R_{\mu\nu}=\frac14\g_{\mu\nu}\R$. This leads to
\beq  \label{eq:85}
\R^{\mu\nu}\R_{\mu\nu}=\frac14\R^2=4(C^\prime)^{-\frac43}r^2\sin\theta 
\eeq 
Eqs.(\ref{eq:84}) and (\ref{eq:85}) are quite different from that of the 
Schwarzschild metric where $\R$ and $\R^{\mu\nu}\R_{\mu\nu}$ are zero everywhere 
including the neighborhood of the origin. The Jacobian of the transformation from polar coordinates $(t,r,\theta,\phi)$ to the coordinate system $(t,x,y,z)$ is 
$r^2\sin\theta$. Since $\R$ is a scalar density of weight $+\frac12$ and $\R^{\mu\nu}
\R_{\mu\nu}$ is a scalar density of weight $+1$, then in the $(t,x,y,z)$ coordinate system we have $\R=-(C^\prime)^{\frac23}$ and $R^{\mu\nu}R_{\mu\nu}=\frac14
(C^\prime)^{-\frac43}$. That is $\R$ is constant and negative while $\R^{\mu\nu}\R_{\mu\nu}$ is constant and positive. To specify the nature of the singularity it is necessary that the scalar density $\R^{\mu\nu\rho\lambda}\R_{\mu\nu\rho\lambda}$ to be calculated too. 
The nonzero components of Riemann tensor are
\beqa  
\R^t_{rtr}&=&(\frac38\frac{B^{\prime\prime}}{B}-\frac{A^{\prime\prime}}{8A}
+\frac{3}{16}(\frac{A^\prime}{A})^2-\frac{3}{16}(\frac{B^\prime}{B})^2-\frac14\frac{A^\prime B^\prime}{AB})\label{eq:86}\\
\R^t_{\theta t\theta}&=&\frac{r^2}{A}(\frac1{64}(\frac{A^\prime}{A})^2-\frac3{64}
(\frac{B^\prime}{B})^2-\frac1{32}\frac{A^\prime B^\prime}{AB}+\frac38\frac{B^\prime}{rB}-\frac{A^\prime}{8rA})\label{eq:87}\\
\R^t_{\phi t\phi}&=&\sin^2\theta\R^t_{\theta t\theta} \label{eq:88}\\
\R^r_{\theta r\theta}&=&\frac{r^2}{A}(-\frac{A^{\prime\prime}}{8A}-\frac{B^{\prime\prime}}{8B}
+\frac3{16}(\frac{A^\prime}{A})^2+\frac18(\frac{B^\prime}{B})^2+\frac1{16}
\frac{A^\prime B^\prime}{AB}-\frac{5}{8r}\frac{A^\prime}{A}-\frac1{8r}\frac{B^\prime}{B}) \label{eq:89}\\
\R^r_{\phi r\phi}&=&\sin^\theta\R^r_{\theta r\theta} \label{eq:90}\\
\R^\theta_{\phi\theta\phi}&=&\frac{r^2}{A}\sin^2\theta(-1+\frac1{r^2}+\frac1{64}(\frac{A^\prime}{A})^2+\frac1{64}(\frac{B^\prime}{B})^2+\frac1{32}\frac{A^\prime B^\prime}{AB}-\frac{A^\prime}{4rA}-\frac1{4r}\frac{B^\prime}{B}) \label{eq:91}
\eeqa
Using Eqs.(\ref{eq:86})-(\ref{eq:91}) we have
\beqa
\R^{\mu\nu\rho\lambda}\R_{\mu\nu\rho\lambda}= 
& & \frac{2B^{\frac12}r^2\sin\theta}{A^{\frac32}}([\frac38
\frac{B^{\prime\prime}}{B}-
\frac{A^{\prime\prime}}{8A}+\frac3{16}(\frac{A^\prime}{A})^2-
\frac3{16}
(\frac{B^\prime}{B})^2-\frac14
\frac{A^\prime B^\prime}{AB}]^2 \n
& & +2[\frac1{64}(\frac{A^\prime}{A})^2-\frac3{64}(\frac{B^\prime}{B})^2-\frac1{32}
\frac{A^\prime B^\prime}{AB}+\frac3{8r}\frac{B^\prime}{B}-\frac{A^\prime}{8rA}]^2 \n
& &+2[-\frac{A^{\prime\prime}}{8A}-\frac{B^{\prime\prime}}{8B}
+\frac3{16}(\frac{A^\prime}{A})^2+\frac18(\frac{B^\prime}{B})^2+\frac
1{16}\frac{A^\prime B^\prime}{AB}-\frac5{8r}\frac{A^\prime}{A}
-\frac1{8r}\frac{B^\prime}{B}]^2 \n
& & +[-1+\frac1{r^2}+\frac1{64}(\frac{A^\prime}{A})^2+
\frac1{64}(\frac{B^\prime}{B})^2+\frac1{32}
\frac{A^\prime B^\prime}{AB}-\frac{A^\prime}{4rA}-\frac{B^\prime}{4rB}]^2) \label{eq:92}
\eeqa
Now inserting the asymptotic forms (\ref{eq:80})-(\ref{eq:83}) for 
$A$ and $B$ into Eq.(\ref{eq:92})yields
\beq \label{eq:93}
\R^{\mu\nu\rho\lambda}\R_{\mu\nu\rho\lambda}\propto\;\frac{\sin\theta}{r^6} \;\;\; 
for \;\;C^\prime>(2GM)^3
\eeq
and
\beq \label{eq:94}
\R^{\mu\nu\rho\lambda}\R_{\mu\nu\rho\lambda}\propto\;\frac{\sin\theta}{r^4} \;\;\;
for  \;\;C^\prime=(2GM)^3
\eeq
$\R^{\mu\nu\rho\lambda}\R_{\mu\nu\rho\lambda}$ is a scalar  density of weight $+1$,
so in a $(t,x,y,z)$ coordinate system Eq.(\ref{eq:93}) and Eq.(\ref{eq:94}) are
proportional to $\frac1{r^8}$ and $\frac1{r^6}$ respectively. This is enough to convince us that $r=0$ represent an actual singularity. 
Nonexistence of event horizon in small distances is a 
novel feature of this work. It is not hard to imagine that taking $\Lambda<0$
which is the counter part of the deSitter spacetime will show an event
horizon at cosmological distances of the order $|\Lambda|^{-\frac12}$. It is worth the alternative version in the 
presence of matter to be investigated.   

\acknowledgments{
A.M.A wants to appreciate supports of research council at university of Tehran.}


\begin{thebibliography}{99}
 
\bibitem{Ein1919} A.Einstein, Sitzungsher.d.Preuss Akad.d.Wissench,Pt.1,433(1919) 
                  . English translation in H.A. Lorentz,
                  A. Einstein etal, The principle of relativity, Dover,New                             Yor(1952).            
           
\bibitem{Fink71} J.Anderson,D.Finkelstein,Amer.J.Phys.{\bf 39/8},901,(1971).
           
\bibitem{Fink01} D.Finkelstein etal.,J.Math.phys.42(1)
                 340-346,Jan 2001.           
           
\bibitem{Bij} J.J.Van der Bij et.al,Physica {\bf{A}}116,367(1982).

\bibitem{Wein89} S.Weinberg,Rev.Mod.Phys.61:1-23,1989.

\bibitem{Alv} E.Alvarez, JHEP 03(2005)002.
     
             
                        
\end{thebibliography}
\end{document}